\documentclass[11pt]{article}
\usepackage[utf8]{inputenc}
\usepackage[OT4]{fontenc}
\usepackage{amssymb}
\usepackage{longtable,lscape}

\begin{document}
\title{New aspects of symmetry of elementary cellular automata}

\author{Małgorzata J. Krawczyk\\
Faculty of Physics and Applied Computer Science,\\
AGH University of Science and Technology,\\
al. Mickiewicza 30, 30-059 Cracow, Poland\\
gos@fatcat.ftj.agh.edu.pl
}
\maketitle
\begin{abstract}
We present a new classification of elementary cellular automata. It is based on the structure of the network of states, connected with the transitions between them; the latter are determined by the automaton rule. Recently an algorithm has been proposed to compress the network of states (M. J. Krawczyk, Physica A 390 (2011) 2181). In this algorithm, states are grouped into classes, according to the local symmetry of the network. In the new classification, an automaton is described by the number of classes $\#(N)$ as dependent on the system size $N$. In most cases, the results reflect the known classification into $88$ groups. However, the function $\#(N)$ also appears to be the same for some rules which have not been grouped together yet. In this way, the automaton $23$ is equivalent to $232$, $77$ to $178$, $105$ to $150$, the pair $(43, 113)$ to the pair $(142, 212)$ and the group $(12, 68, 207, 221)$ to the group $(34, 48, 187, 243)$. Furthermore, automata $51$, $204$, the pair $(15, 85)$ and the pair $(
170,240)$ are all mutually equivalent. Results are also presented on the structure of networks of states.
\end{abstract}

\section{Introduction}
Attempts of cellular automata classification were made by several authors. Proposed approaches are based, either on observations of the automata behaviour, or properties of the rules themselves. The most famous classification of the elementary cellular automata CA belongs to the first category, and was introduced by Wolfram \cite{wolf}. The classification regarded the character of automata evolution and led to the division of the whole set of elementary rules into four classes. The first group includes rules which applied to a given sequence, cause removal of any randomness, and as a consequence lead to a stable state. The rules included into the second class lead to a stable or oscillating character of evolution. The third class covers rules which show with pseudo-random or chaotic behaviour. The fourth class is especially interesting as it covers rules which lead to complex evolution. The classification makes it possible to see differences in behaviour of the system observed during the evolution ruled by 
the
particular automata.\\

Besides the Wolfram's classification there are also different propositions. One of them was introduced by Li at al. \cite{lipa}. The classification is based on the observation of the asymptotic behaviour of the elementary CA, and, as a result, six types are distinguished. CA for which evolution leads to: a) spatially homogeneous fixed point, b) spatially inhomogeneous fixed point or a uniform shift of a fixed pattern, c) periodic behaviour or shifted periodic behaviour, d) locally chaotic behaviour, e) chaotic behaviour or f) complex behaviour.
Another classification is based on the concept of equicontinuity which originally was introduced by Gilman \cite{gilm} and then extended by Kurka \cite{kurk}, where an equicontinuity set contains a set of its Lyapunov stable points. In this scheme, similarly to Wolfram's classification, each CA can be included into one of the four categories: equicontinuous CA, CA with some equicontinuous points, sensitive but not positively expansive and positively expansive. It is not clear \textit{a priori} to which category a particular automaton belongs, but for the elementary CA this problem can be solved \cite{schu}. A review of the classification methods, along with a discussion of problems that arise when attempting classification of automata, is presented by Sutner in \cite{sutn}. Analysis of cellular automata properties is an important problem as they have a lot of applications in modelling of physical processes, such as diffusion, phase transition, wave propagations or road traffic analysis \cite{dro}.

In this paper we propose a method of class identification which is based on the similarities observed in patterns of connections between possible states of the analysed system. Here we say that two states are connected if one state is obtained from another one, according to a given automaton rule. The presented method takes into account the symmetry observed in the state space, and it allows us to indicate equivalent - according to symmetry - rules. The method is exhaustive, and is based on scrupulous analysis performed for sequences of lengths up to $N=19$ (in some cases longer if the character of results does not allow conclusions to be drawn earlier).\\

The paper is organised as follows. In the next section the analysed system is presented. In the section after that our procedure for identification of classes is described. Section \ref{results} is devoted to discussion about the obtained results, and the last section provides their summary.

\section{Analysed system and symmetry of rules}

We analyse a one-dimensional cellular automaton, with two possible states of each cell, $0$ and $1$. The rule of the change of a state of a cell is determined on the basis of the state of the cell itself and its two nearest neighbours. Namely, the rule is a chain $(A,B,...,H)$ of outcomes of the states of these three cells, in a decreasing order $(111,110,...,000)$. Such a definition leads to $2^8=256$ different rules of the evolution process, so called elementary cellular automata. It was however shown that the number of unique rules is lower, as some of the rules are equivalent \cite{wolf1}. The rules which are equivalent to a given one are obtained as its conjugation, reflection and combined operation of conjugation and reflection (see Tab.\ref{equiv}). As a result, the set of all $256$ elementary automata is reduced into $88$ unique rules, or in other words, to $88$ groups of equivalent rules. The same result was recently obtained in \cite{mach} by an analysis of the symmetry of the rules. It may happen 
however, that after any change, the obtained rule is the same as at the beginning. Because of that, symmetry groups do not always contain four rules. To be precise, there are $44$ groups which contain $4$ rules each, $36$ groups which contain $2$ rules each, and a remaining $8$ rules which form one-element groups. Illustrative examples are presented in Tab.\ref{examp}.\\

\begin{table}
\begin{center}
\begin{tabular}{{l|}*{8}{c}}
\textit{rule}&\textbf{$A$}&\textbf{$B$}&\textbf{$C$}&\textbf{$D$}&\textbf{$E$}&\textbf{$F$}&\textbf{$G$}&\textbf{$H$}\\\hline
\textit{conjugation}&\textbf{$\bar{H}$}&\textbf{$\bar{G}$}&\textbf{$\bar{F}$}&\textbf{$\bar{E}$}&\textbf{$\bar{D}$}&\textbf{$\bar{C}$}&\textbf{$\bar{B}$}&\textbf{$\bar{A}$}\\\hline
\\[-.4cm]
\textit{reflection}&\textbf{$A$}&\textbf{$E$}&\textbf{$C$}&\textbf{$G$}&\textbf{$B$}&\textbf{$F$}&\textbf{$D$}&\textbf{$H$}\\\hline
\\[-.4cm]
\textit{conjugation+reflection}&\textbf{$\bar{H}$}&\textbf{$\bar{D}$}&\textbf{$\bar{F}$}&\textbf{$\bar{B}$}&\textbf{$\bar{G}$}&\textbf{$\bar{C}$}&\textbf{$\bar{E}$}&\textbf{$\bar{A}$}\\
\end{tabular}
\end{center}
\caption{Rule equivalence. The bar means negation, i.e. $\bar{0}=1$ and $\bar{1}=0$. }
\label{equiv}
\end{table}

\begin{table}
a)\\[-1cm]
\begin{center}
\begin{tabular}{{l|}*{8}{c}|l}
\textit{rule}&\textbf{$0$}&\textbf{$0$}&\textbf{$0$}&\textbf{$0$}&\textbf{$0$}&\textbf{$1$}&\textbf{$1$}&\textbf{$1$}&\textit{No 7}\\\hline
\textit{conjugation}&\textbf{$0$}&\textbf{$0$}&\textbf{$0$}&\textbf{$1$}&\textbf{$1$}&\textbf{$1$}&\textbf{$1$}&\textbf{$1$}&\textit{No 31}\\\hline
\textit{reflection}&\textbf{$0$}&\textbf{$0$}&\textbf{$0$}&\textbf{$1$}&\textbf{$0$}&\textbf{$1$}&\textbf{$0$}&\textbf{$1$}&\textit{No 21}\\\hline
\textit{conjugation+reflection}&\textbf{$0$}&\textbf{$1$}&\textbf{$0$}&\textbf{$1$}&\textbf{$0$}&\textbf{$1$}&\textbf{$1$}&\textbf{$1$}&\textit{No 87}\\
\end{tabular}
\end{center}
b)\\[-1cm]
\begin{center}
\begin{tabular}{{l|}*{8}{c}|l}
\textit{rule}&\textbf{$1$}&\textbf{$0$}&\textbf{$0$}&\textbf{$0$}&\textbf{$0$}&\textbf{$1$}&\textbf{$0$}&\textbf{$0$}&\textit{No 132}\\\hline
\textit{conjugation}&\textbf{$1$}&\textbf{$1$}&\textbf{$0$}&\textbf{$1$}&\textbf{$1$}&\textbf{$1$}&\textbf{$1$}&\textbf{$0$}&\textit{No 222}\\\hline
\textit{reflection}&\textbf{$1$}&\textbf{$0$}&\textbf{$0$}&\textbf{$0$}&\textbf{$0$}&\textbf{$1$}&\textbf{$0$}&\textbf{$0$}&\textit{No 132}\\\hline
\textit{conjugation+reflection}&\textbf{$1$}&\textbf{$1$}&\textbf{$0$}&\textbf{$1$}&\textbf{$1$}&\textbf{$1$}&\textbf{$1$}&\textbf{$0$}&\textit{No 222}\\
\end{tabular}
\end{center}
c)\\[-1cm]
\begin{center}
\begin{tabular}{{l|}*{8}{c}|l}
\textit{rule}&\textbf{$0$}&\textbf{$1$}&\textbf{$0$}&\textbf{$0$}&\textbf{$1$}&\textbf{$1$}&\textbf{$0$}&\textbf{$1$}&\textit{No 77}\\\hline
\textit{conjugation}&\textbf{$0$}&\textbf{$1$}&\textbf{$0$}&\textbf{$0$}&\textbf{$1$}&\textbf{$1$}&\textbf{$0$}&\textbf{$1$}&\textit{No 77}\\\hline
\textit{reflection}&\textbf{$0$}&\textbf{$1$}&\textbf{$0$}&\textbf{$0$}&\textbf{$1$}&\textbf{$1$}&\textbf{$0$}&\textbf{$1$}&\textit{No 77}\\\hline
\textit{conjugation+reflection}&\textbf{$0$}&\textbf{$1$}&\textbf{$0$}&\textbf{$0$}&\textbf{$1$}&\textbf{$1$}&\textbf{$0$}&\textbf{$1$}&\textit{No 77}\\
\end{tabular}
\end{center}
\caption{Examples of groups of equivalent rules: a) Group contains $4$ different rules No $7, 21, 31$ and $87$, b) Group contains $2$ different rules No $132$ and $222$, c) Group contains rule No $77$}
\label{examp}
\end{table}

Consider a system consisting of all possible sequences of zeros and ones of a specified length $N$. Each sequence can be converted to an exact single sequence determined by the automaton rule which is currently being considered. Each sequence may, however, be obtained by the transformation of different numbers of sequences. The exact number of states (sequences) leading to a given state depends on the automaton rule and the system size $N$. The state space considered can be represented as a directed graph, where $2^N$ nodes are identified with the possible states of the system. As for a given automaton, each state can be transformed to only one other state the out-degree of all nodes is equal to $1$. However, the in-degree is different for different nodes.

\section{Classes}
Let us consider a given automaton rule. Analysis of the pattern of ties in the obtained graph allows us to indicate nodes which are similar, which means that the structure of connections of some nodes is the same. Because of that, nodes can be divided into classes, and each class contains nodes which are similar. As a result, the state space formed by all possible classes is smaller than the initial network of states. The rate of the reduction depends on the system size.\\

The concept of the reduction of the system size is based on the observation that, often in real systems, some kind of symmetry can be uncovered. If it is the case, it is possible to specify groups of states of the analysed system which manifest similar properties. When one thinks about the system like a graph, the mentioned similarities concern primarily nodes degrees. However, this quantity by itself does not reflect the node properties. If two nodes have the same degree, but the degrees of their neighbours are different, the nodes cannot be classified as being similar. If weights of particular links in a graph are not equal, which is not the case here, one must also distinguish identical patterns of connections but with different weights.\\

The class is defined as the set of nodes which have the same number of neighbours which belong to the same classes. The method of classes (groups of similar nodes) identification was described in detail in our previous papers \cite{mk1,mk2,mk3}. However, in the case of cellular automata the procedure is slightly different from that previously used. The underlying reason for the difference is that now the graph of transitions between states is directed. Therefore, the class of states is determined both by the list of states which for a particular automaton lead to a given state, and the one-element list, which preserve information about the state which is obtained from the one actually being considered. The parameter which differentiate two nodes is their degree, so, at the very beginning of our procedure, all nodes which have the same degree are labelled with the same symbol. With this done, the next step is to check if the lists of symbols of neighbours of all nodes which have assigned a given symbol are 
exactly the same. If not, appropriate elements have to be distinguished: for each class where this ambiguity is observed, as many separate classes must be introduced as it is necessary to ensure that for any two states, which belong to a given class, the full agreement for both lists - states which lead to a given state and the state which are obtained from it - is obtained.  In general, the obtained network of classes is weighted, as more than one of the neighbours of a given node in the network of states may belong to the same class. In a further analysis of the obtained network of classes one must take into account the number of states which belong to particular classes. The procedure usually leads to a significant reduction of the size of the system.\\

It must be emphasised, that for each system size $N$, the class structure obtained for the rules known to be equivalent \cite{wolf1,mach} should be exactly the same. An interesting question is if it is also true for the rules which belong to different symmetry groups, but equivalent according to the function $\#(N)$.

\section{Results}
\label{results}
\subsection{Number of classes in a function of system size for particular automata}
The number of classes for particular automata changes with the system size~$N$. We can indicate a few different types of this dependence:
\begin{itemize}
\item
\textit{automata No $0$ and $255$}\\
the number of classes is constant regardless of the system size and equals $2$; one class contains states composed of zeros or ones, respectively, and the second class contains all remaining possible sequences. All of them are transformed in one case to the state composed of zeros, and in the other to the state composed of ones. In other words, in this case we obtain the star graph
\item
\textit{automata No $15, 51, 85, 170, 204$ and $240$}\\
the second characteristic group of rules contains those for which just one class is obtained, irrespective of the system size; the rules cause that both \textit{in} and \textit{out} degree of each state is equal to $1$; for the rules which belong to this group a new state of each cell is obtained as follows: for rule No $204$ a new state is equal to the previous one, for rule No $51$ it is negation of the previous state; for the four remaining rules a new state is determined on the basis of the previous state of the one of the nearest neighbours, and it is the state currently assigned to the appropriate neighbour or its negation, namely it is: for rule No $170$ the state of the right neighbour, for rule No $85$ the negation of the right neighbour, for rule No $240$ the state of the left neighbour, and for rule No $15$ the negation of the left neighbour
\item
\textit{automata No $105$ and $150$}\\
the number of classes in the case of these two rules form a sequence presented in Tab.\ref{seq105}; during evolution in accordance with rule No $105$, if the number of ones in the triple of cells (a given cell and its two nearest neighbours) is even - $0$ or $2$ - irrespective of the previous state of a given cell, it is set to $1$, if it is odd the value is set to $0$. In the case of rule No $150$ the situation is opposite - the new value is equal to $0$ if the number of ones is even, and to $1$ if it is odd
\begin{table}
\begin{center}
\begin{tabular}{c*{19}{|c}}
\textbf{$N$}&3&4&5&6&7&8&9&10&11&12&13&14&15&16&17&18&19&20&21\\\hline
\textbf{$\#$}&2&1&1&3&1&1&2&1&1&5&1&1&2&1&1&3&1&1&2\\
\end{tabular}
\end{center}
\caption{Number of classes found for the automata No~$105$ and $150$ in a function of the system size $N$.}
\label{seq105}
\end{table}
\item
\textit{automata No $154, 166, 180$ and $210$}\\
in this case, if $N$ is an odd number, the number of classes equals $1$, whereas if it is even the number of classes form a sequence presented in Tab.\ref{seq154}; it may be noted that for every second even value (indivisible by $4$) the number of classes is equal to $5$; for all four rules classified here we observe the four configurations of a given cell state and its neighbours lead to the change of the cell state, and in particular if all three states are the same, the state of the considered cell does not change
\begin{table}
\begin{center}
\begin{tabular}{c*{9}{|c}}
\textbf{$N$}&4&6&8&10&12&14&16&18&20\\\hline
\textbf{$\#$}&4&5&6&5&7&5&10&5&7\\
\end{tabular}
\end{center}
\caption{Number of classes found for the automata No~$154, 166, 180$ and $210$ in a function of its even $N$, in $N$ in odd number of classes equals $1$.}
\label{seq154}
\end{table}
\item
\textit{automata No $45, 75, 89$ and $101$}\\
here, similarly to the previous case, if $N$ is odd the number of classes is equal to $1$, but if $N$ is even the number of classes increases exponentially with $N$ for $N\gtrapprox 10$; the rules classified to this category are a mirror image of the rules described above. Because of that, observations about the number of changed states remain the same, but in this case, if states of all cells in the triple are the same, the state of the cell in the middle is flipped
\item
\textit{automata No $60, 102, 153$ and $195$}\\
for those four automata if the system size is an odd number, the number of classes equals $2$, and if the system size is even the number of classes changes with $N$ in the way presented in Tab.\ref{seq60}; for every second even value (indivisible by $4$) the number of classes equals $3$; all rules classified here are symmetric with respect to its center, and also in this case four configurations of states in the triple of cells lead to change and to maintain the current cell state
\begin{table}
\begin{center}
\begin{tabular}{c*{9}{|c}}
\textbf{$N$}&4&6&8&10&12&14&16&18&20\\\hline
\textbf{$\#$}&5&3&9&3&5&3&17&3&5\\
\end{tabular}
\end{center}
\caption{Number of classes found for automata No~$60, 102, 153$ and $195$ in a function of its even $N$, in $N$ in odd number of classes equals $2$.}
\label{seq60}
\end{table}
\item
\textit{automata No $90$ and $165$}\\
the pattern of change of the number of classes in the case of those two automata is particularly interesting, more because the rule produce in the evolution the Sierpinski triangle fractal \cite{wolf}; the number of classes obtained in this case is presented in Fig.\ref{fig1}; in this case, half of the triples also lead to the modification of the current cell value, and, for rule No $90$, a new value is equal to $1$ if the value of one of the nearest neighbours has the same value as the currently considered cell, otherwise it is equal to $0$; for rule No $165$ the situation is opposite
\begin{figure}
\begin{center}
\includegraphics{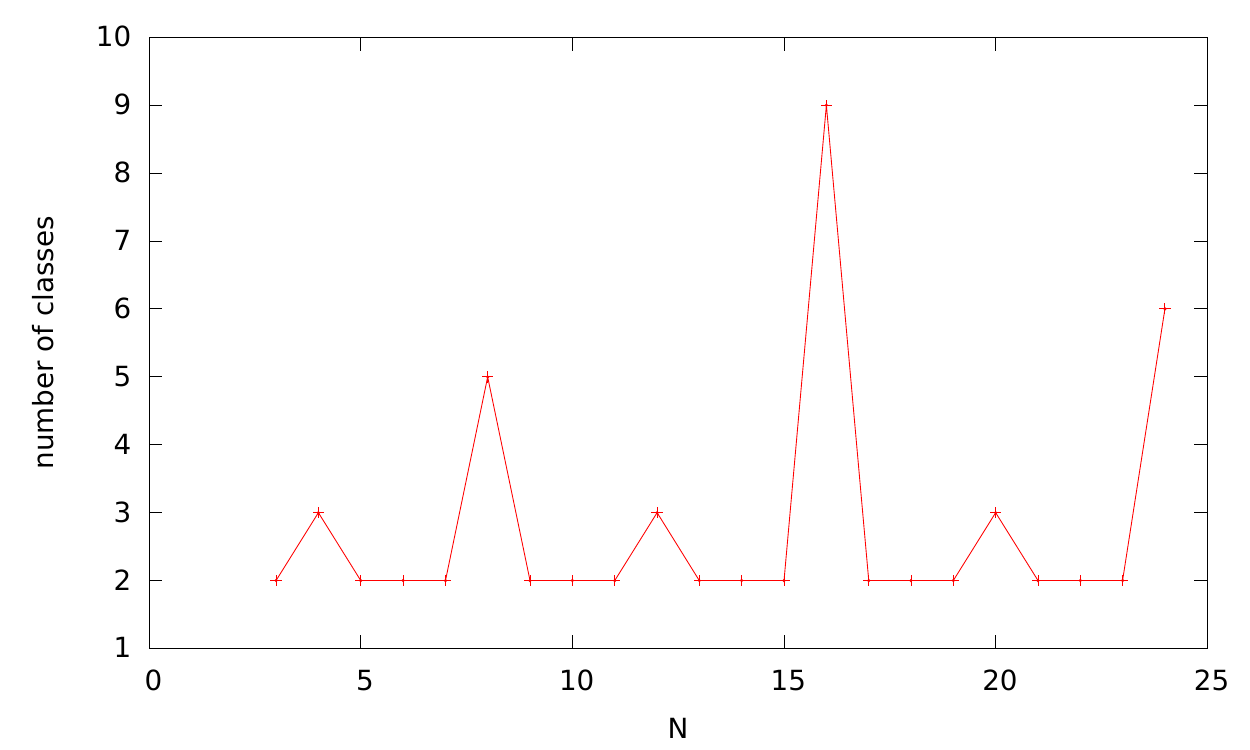}
\caption{Number of classes in a function of the system size for automata No $90$ and $165$.}
\label{fig1}
\end{center}
\end{figure}
\item
remaining automata\\
for most automata, none specified above, the number of classes increases exponentially with the system size, although slower than the number of possible sequences. The number of classes grows at the most as approximately $1.7^N$, while the number of states as $2^N$, where $N$ is the system size. The largest number of classes is observed for the automaton No $30$, which belongs to the third class in the Wolfram classification system, and displays chaotic behaviour \cite{wolf}. Of course the same number of classes was identified for $3$ remaining automata which belong to the same symmetry group (in a Wolfram's sense), it is the automata No $86, 135$ and $149$. The second largest number of classes is observed for the automata No $110$, $124, 137$ and $193$, known to be Turing complete \cite{wolf, cook}. Those automata demonstrate a mixed character of evolution, where regular and irregular behaviour is observed.\\
The reduction of the size of the system for different automata is very different, and changes from one to even four orders of magnitude. The obtained values should be compared with $2^N$, i.e. the number of states. A histogram of the number of classes $n$ for $N=19$ is presented in Fig.\ref{fig2}.

\begin{figure}
\begin{center}
\includegraphics{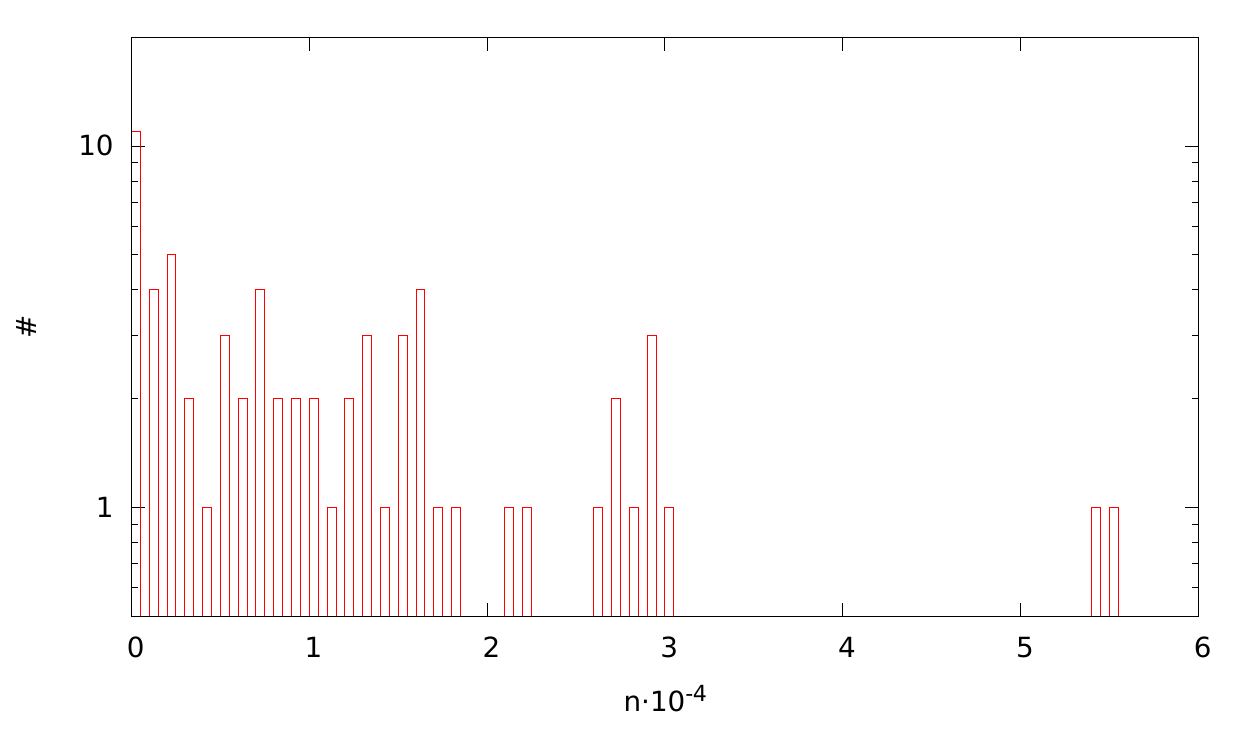}
\caption{Histogram of the number of classes $n$ for $N=19$ in the case of the group of the automata for which exponential increase of the number of classes with the system size is observed. Note that $2^{19}=524288$, then the reduction is by at least one order of magnitude.}
\label{fig2}
\end{center}
\end{figure}

An exponential increase of the number of classes with the system size is very often observed in the whole range of analysed lengths, and only in some cases a deviation from this rule appears for automata shorter than approximately $10$ cells. Among automata for which an exponential increase of the  number of classes is observed, a group for which the number of classes depends on the system size parity is noticeable.\\
\end{itemize}

All but two rules for which atypical (different that exponential) character of the change of the number of classes with the sequence size are classified in \cite{schu} as surjective rules. The rules which do not fulfil this property are automata No $0$ and $255$. What's more, most of them are also classified as being chaotic in accordance with the definition of chaos presented in \cite{dev}. For all those automata value of \textit{z-parameter} \cite{wue} is equal to $1$, which is expected as they are chaotic. However, our classification indicate that for two symmetry groups in the Wolfram sense which are also classified as being chaotic, namely the groups $[30,86, 135,149]$ and $[106, 120,169,225]$, a typical - exponential - increase of the number of classes with the sequence size is observed.

\subsection{Networks of states and classes}
Our results show that the same number of classes can be found for automata which belong to different groups, out of $88$ possibilities; in most cases this happens for a small number of groups for small $N$. Whether the structure of classes is the same or the observed compliance concerns only the number of classes, depends on the system size. However, in some cases the equivalence seems to be independent of $N$. The detailed description is presented in Tab.\ref{longt}. The first column contains $88$ groups of rules, one after another. The following columns present the numbers of classes identified, as dependent on the system size $N$. The number of classes is supplemented with a letter, which makes it possible to distinguish cases for which the number of classes is the same but the lists of neighbouring states (in- and out-states) are different. The same class structure for different groups is observed or not, depending on $N$.\\

For automata for which the same symbol was assigned, the complete consistency of the networks of states and the (reduced) networks of classes is observed. As a criterion of the equivalence of networks of states we take the identity of lists of in- and out-neighbours of all states. In Tab.\ref{longt} it is seen, for example, that the symbol $1A$ for $N=5$ appears $8$ times, while for $N=18$ only $4$ times. The result means that in a previous case for $8$ symmetry groups, namely $(15,85)$, $(45,75,89,101)$, $(51)$, $(105)$, $(150)$, $(154,166,180,210)$, $(170,240)$ and $(204)$, networks of states and classes for $N=5$ is exactly the same. The same is true for $N=18$ for groups $(15,85)$, $(51)$, $(170,240)$ and $(204)$. The case of automata No $51$, $204$ and groups $(170, 240)$ and $(15, 85)$ should be considered separately. In this case, both networks - states and classes - are the same ($1A$) regardless of the system size $N$. This is because these automata are reversible: the identity and negation, the 
left-
shift and the right-shift and their negation. There, states take part in long (e.g. left-shift) or short (identity) loops. In principle, there is no difference between states. Then, each state belongs to the same class, and there is only one class. Hence, each state is produced by a state from the same class and also gives a state from the same class. Hence, the reduced network is exactly the same.\\

On the other hand, it is worth noting that in the case of a few groups, the pattern of the change of the number of classes with the system size is exactly the same. This is true in the case of automata No $23$ and $232$, $77$ and $178$, $105$ and $150$, the pair $(43, 113)$ and the pair $(142, 212)$, the group $(12, 68, 207, 221)$ and the group $(34, 48, 187, 243)$. A conjecture appears, and this equivalence is an indication of some symmetry of the rules, different than the conjugation and the reflection. Apart from operations which allow us to indicate equivalent rules, one can also consider negation and reversion of rules Tab.\ref{equiv1}. Those operations lead to sequences, which belong to the same symmetry group, usually different than the groups to which the original sequence belongs. For $16$ elementary automata, the sequence obtained as a negation and as a reversion is exactly the same. Those automata form $8$ one-element and $4$ two-elements symmetry groups Tab.\ref{equiv2}. The group which contain 
the transformed sequence is joined with the group which contain the original one. We also observe that two four-elements groups which contain sequences obtained in accordance with the negation and reversion operations, are joined (Tab.\ref{equiv3}.) because of the same pattern of change of the number of classes.\\

If we take into account the pattern of change of the number of classes, we see that the number of symmetry groups is equal to $80$ (the number of different rows in Tab.\ref{longt}). However, it is possible that this equivalence is broken for larger systems.\\

\begin{table}
\begin{center}
\begin{tabular}{{l|}*{8}{c}}
\textit{rule}&\textbf{$A$}&\textbf{$B$}&\textbf{$C$}&\textbf{$D$}&\textbf{$E$}&\textbf{$F$}&\textbf{$G$}&\textbf{$H$}\\\hline
\\[-.4cm]
\textit{negation}&\textbf{$\bar{A}$}&\textbf{$\bar{B}$}&\textbf{$\bar{C}$}&\textbf{$\bar{D}$}&\textbf{$\bar{E}$}&\textbf{$\bar{F}$}&\textbf{$\bar{G}$}&\textbf{$\bar{H}$}\\\hline
\textit{reversion}&\textbf{$H$}&\textbf{$G$}&\textbf{$F$}&\textbf{$E$}&\textbf{$D$}&\textbf{$C$}&\textbf{$B$}&\textbf{$A$}\\\hline
\end{tabular}
\end{center}
\caption{Rule symmetry. The bar means negation, i.e. $\bar{0}=1$ and $\bar{1}=0$. }
\label{equiv1}
\end{table}

\begin{table}
\begin{center}
\begin{tabular}{l|*{9}{c}}
No&\multicolumn{8}{c}{rule}\\\hline
23&0&0&0&1&0&1&1&1\\\hline
232&1&1&1&0&1&0&0&0\\\hline\hline
77&0&1&0&0&1&1&0&1\\\hline
178&1&0&1&1&0&0&1&0\\\hline\hline
105&0&1&1&0&1&0&0&1\\\hline
150&1&0&0&1&0&1&1&0\\\hline\hline
51&0&0&1&1&0&0&1&1\\\hline
204&1&1&0&0&1&1&0&0\\\hline\hline
15&0&0&0&0&1&1&1&1\\
85&0&1&0&1&0&1&0&1\\\hline
170&1&0&1&0&1&0&1&0\\
240&1&1&1&1&0&0&0&0\\\hline\hline
43&0&0&1&0&1&0&1&1\\
113&0&1&1&1&0&0&0&1\\\hline
142&1&0&0&0&1&1&1&0\\
212&1&1&0&1&0&1&0&0
\end{tabular}
\end{center}
\caption{Equivalence of the selected symmetry groups. Single horizontal lines separate different symmetry groups, and double lines separate groups which are joined.}
\label{equiv2}
\end{table}

\begin{table}
\begin{center}
\begin{tabular}{l|*{9}{c}}
No&\multicolumn{8}{c}{rule}\\\hline
12&0&0&0&0&1&1&0&0\\
68&0&1&0&0&0&1&0&0\\
207&1&1&0&0&1&1&1&1\\
221&1&1&0&1&1&1&0&1\\\hline
34&0&0&1&0&0&0&1&0\\
48&0&0&1&1&0&0&0&0\\
187&1&0&1&1&1&0&1&1\\
243&1&1&1&1&0&0&1&1
\end{tabular}
\end{center}
\caption{Equivalence of the selected symmetry groups. The single horizontal line separates different symmetry groups.}
\label{equiv3}
\end{table}

\subsection{Number of groups in a function of system size}

It is obvious that the structure of our classes is exactly the same for all rules which belong to each from among $88$ unique symmetry groups \cite{wolf1,mach}. However, our class identification procedure allows us to take into account, not only the symmetry which characterizes the automata rules \cite{wolf1,mach}, but also the symmetry which manifests in the structure of transitions between particular possible sequences of a given length. Here, a group consists of automata for which the number of classes $\#(N)$ and the network of states as dependent on the system size $N$ is the same. This is the basis for our classification. As a result, the number of groups for any system size is less than $88$. The number of groups in a function of the system size is presented in Tab.\ref{equiv4}

\begin{table}
\begin{center}
\begin{tabular}{c*{15}{|c}}
\textbf{$N$}&5&6&7&8&9&10&11&12&13&14&15&16&17&18&19\\\hline
\textbf{$\#$}&60&76&70&77&72&77&71&79&71&78&72&78&71&79&71\\
\end{tabular}
\end{center}
\caption{Number of groups as dependent on the system size $N$.}
\label{equiv4}
\end{table}

\subsection{Cycles}

If the number of classes is equal to $1$, both \textit{in} and \textit{out} degree of each node equals $1$. It means that there is a sequence of configurations in which, starting from a given configuration, passing through other possible configurations of the system, we are able to return to the initial state. We examine the length of such cycles for the automata for which one class was obtained. As a result, histograms of the cycle lengths for different automaton lengths were constructed. In most cases the length of the cycle is an integer multiple of the system size.\\

An example is presented in Tab.\ref{A45}, where the result of the number of cycles of a given length for the rule No $45$ is shown for $N=15$. Symbols used in the table specify: $L_c$ is a cycle length, $n_c$ is the number of cycles of a given length and $m$ shows how many times $N$ is housed completely in $L_c$. The example presented is representative and it shows all possible situations which may be observed. First of all we can observe the situation where the length of the cycle is very short, in the example presented it is $1, 2$ or $3$ configurations. The length of the cycle equal to $1$ means that the given configuration passes into itself. Then, we can observe a few cases where the length of the cycle is a multiple of the configuration length, in the case of the automaton No $45$ for $N=15$, it is the case in $6$ cases. In most cases we also observe one cycle which appears as many times as the length of the configuration does (in Tab.\ref{A45} marked as $\ast$). Finally, lengths of cycles which cannot 
be easily interpreted may also arise (in Tab.\ref{A45} marked as $\diamond$). As we have seen above, a cycle often is equivalent to one class.\\
\begin{table}
\begin{center}
\begin{tabular}{c*{11}{|c}}
\textbf{$L_c$}&1&2&3&30&32&60&120&2340&2820&4920&6820\\\hline
\textbf{$n_c$}&3&1&1&2&15&21&3&1&1&1&3\\\hline
\textbf{$m$}&-&-&-&2&$\ast$&4&8&156&188&328&$\diamond$\\
\end{tabular}
\end{center}
\caption{Lengths of cycles found for the automata No~$45$ on lengths $N=15$ (number of classes equals $1$). Here, $L_c$ is a cycle length, $n_c$ is the number of cycles of a given length and $m$ shows how many times $N$ is housed completely in $L_c$}
\label{A45}
\end{table}

\section{Discussion}
We present a new method of cellular automata classification which allows us to differentiate rules in respect of the character of connections between all possible states of the system of a given size. We show that different types of character of the change of the number of classes with the system size is observed for the elementary CA. For most of them, an exponential increase is observed, but we also observe some number of rules which show different kinds of behaviour. Namely, besides the obvious case of automata No $0$ and $255$, a non-trivial dependence of the number of classes in a function of the system size is observed for automata No $15, 45, 60, 90, 105$ and $ 154$ (and their equivalent rules). In most cases, the class identification allows a significant reduction of the system size. The proposed method indicates that the number of symmetry groups is equal to $80$, as for some groups indicated by Wolfram, the same pattern of the change of the number of classes for different system
size is observed. The proposed method also makes it possible to show that number of symmetry groups changes with the system size.\\
The class reduction speeds up the calculation of the stationary probability distribution of particular configurations \cite{mk3}. In particular, all „Garden of Eden” configurations have the probability zero, and all configurations which form a cycle have the same probabilities. In our method, it appears that all states which belong to a given class have the same probability. Hence, once we are able to calculate the probability of a given class we also know the probability of each state which belongs to this class. As a consequence, any analysis which concerns stationary properties of the analysed system may be performed on a reduced, smaller representation of the system. The method was described in \cite{mk3}. Although the results presented here are obtained for the elementary cellular automata, our method can be applied to any family of automata, irrespective of its rules and size of its neighbourhood. It should be also emphasised that the method is general, and it can be applied to any system for which we are able to define a state space and an elementary process which transforms one state into another. As our method is numerical and not analytical, one should be cautious in predicting an asymptotic behaviour of the system. What we are able to conclude, is the behaviour of the system in the range of finite N.

{\bf Acknowledgement:}
The author is grateful to Krzysztof~Kułakowski for his critical reading of the manuscript and helpful discussion. The work was partially supported by the Polish Ministry of Science and Higher Education and its grants for Scientific Research, and by PL-Grid Infrastructure.

\begin{landscape}
{\footnotesize{
\begin{longtable}{c*{15}{|c}}
\caption{Number of classes for different $N$ for symmetry groups}\\
\textbf{$No\downarrow\vert N\rightarrow$}&$5$&$6$&$7$&$8$&$9$&$10$&$11$&$12$&$13$&$14$&$15$&$16$&$17$&$18$&$19$\\\hline
\hline
\endfirsthead
\multicolumn{16}{c}%
{\tablename\ \thetable\ -- \textit{Continued from previous page}} \\
\textbf{$No\downarrow\vert N\rightarrow$}&$5$&$6$&$7$&$8$&$9$&$10$&$11$&$12$&$13$&$14$&$15$&$16$&$17$&$18$&$19$\\\hline
\hline
\endhead
\hline \multicolumn{16}{r}{\textit{Continued on next page}} \\
\endfoot
\endlastfoot
\textbf{$0, 255$} &$2A$ &$2C$ &$2E$ &$2F$ &$2G$ &$2H$ &$2I$ &$2J$ &$2K$ &$2L$ &$2M$ &$2N$ &$2O$ &$2P$ &$2Q$ \\\hline
\textbf{$1, 127$} &$4A$ &$7M$ &$9K$ &$12H$ &$14N$ &$19H$ &$24H$ &$31G$ &$43C$ &$60C$ &$81C$ &$112A$ &$149A$ &$200B$ &$262B$ \\\hline
\textbf{$2, 16, 191, 247$} &$4B$ &$5G$ &$6W$ &$6Z$ &$9S$ &$12K$ &$12M$ &$17O$ &$20I$ &$24I$ &$31I$ &$38C$ &$44C$ &$57D$ &$68D$ \\\hline
\textbf{$3, 17, 63, 119$} &$7A$ &$10A$ &$12F$ &$17H$ &$22I$ &$32C$ &$44B$ &$65B$ &$92B$ &$130B$ &$178A$ &$246B$ &$333A$ &$452B$ &$606A$ \\\hline
\textbf{$4, 223$} &$5A$ &$5H$ &$7Q$ &$9N$ &$11W$ &$13S$ &$17M$ &$21F$ &$25F$ &$33H$ &$37D$ &$49E$ &$57C$ &$71D$ &$85A$ \\\hline
\textbf{$5, 95$} &$7A$ &$9A$ &$12F$ &$9O$ &$22I$ &$24G$ &$44B$ &$46C$ &$92B$ &$62B$ &$178A$ &$117C$ &$333A$ &$187A$ &$606A$ \\\hline
\textbf{$6, 20, 159, 215$} &$7B$ &$9B$ &$14H$ &$20D$ &$37A$ &$53A$ &$86B$ &$142A$ &$235A$ &$376A$ &$647A$ &$1052A$ &$1791A$ &$2951A$ &$5051A$ \\\hline
\textbf{$7, 21, 31, 87$} &$8A$ &$14A$ &$19A$ &$33A$ &$48A$ &$83A$ &$127A$ &$218A$ &$341A$ &$600A$ &$990A$ &$1691A$ &$2842A$ &$4849A$ &$8251A$ \\\hline
\textbf{$8, 64, 239, 253$} &$4C$ &$6R$ &$6X$ &$6AA$ &$10N$ &$12L$ &$12N$ &$18F$ &$20J$ &$24J$ &$32F$ &$38D$ &$44D$ &$58B$ &$68E$ \\\hline
\textbf{$9, 65, 111, 125$} &$6A$ &$11A$ &$13K$ &$22A$ &$34B$ &$53B$ &$80A$ &$133A$ &$211A$ &$352A$ &$567A$ &$940A$ &$1559A$ &$2611A$ &$4345A$ \\\hline
\textbf{$10, 80, 175, 245$} &$4D$ &$2D$ &$6Y$ &$11U$ &$12J$ &$6AB$ &$20H$ &$27H$ &$30G$ &$12O$ &$46D$ &$63B$ &$68C$ &$36D$ &$96C$ \\\hline
\textbf{$11, 47, 81, 117$} &$5B$ &$11B$ &$13L$ &$24A$ &$33D$ &$51B$ &$79A$ &$141B$ &$209B$ &$354A$ &$571A$ &$938A$ &$1502A$ &$2457B$ &$3974A$ \\\hline
\textbf{$12, 68, 207, 221$} &$4D$ &$9C$ &$6Y$ &$15H$ &$12J$ &$21D$ &$20H$ &$33F$ &$30G$ &$49B$ &$46D$ &$69D$ &$68C$ &$99B$ &$96C$ \\\hline
\textbf{$13, 69, 79, 93$} &$6B$ &$11C$ &$15A$ &$25A$ &$39A$ &$60B$ &$92A$ &$158A$ &$246A$ &$417A$ &$670A$ &$1107A$ &$1796A$ &$3042A$ &$4980A$ \\\hline
\textbf{$14, 84, 143, 213$} &$7C$ &$11D$ &$15B$ &$26A$ &$37B$ &$66A$ &$99A$ &$161A$ &$262A$ &$464A$ &$716A$ &$1170A$ &$2008A$ &$3480A$ &$5465A$ \\\hline
\textbf{$15, 85$} &$1A$ &$1A$ &$1A$ &$1A$ &$1A$ &$1A$ &$1A$ &$1A$ &$1A$ &$1A$ &$1A$ &$1A$ &$1A$ &$1A$ &$1A$ \\\hline
\textbf{$18, 183$} &$4E$ &$8N$ &$11P$ &$17I$ &$24F$ &$36C$ &$52F$ &$83B$ &$117B$ &$189A$ &$295A$ &$467A$ &$732A$ &$1201A$ &$1942A$ \\\hline
\textbf{$19, 55$} &$8B$ &$10B$ &$14I$ &$19E$ &$25D$ &$34E$ &$46B$ &$66B$ &$92C$ &$132C$ &$183B$ &$258A$ &$353A$ &$491A$ &$665A$ \\\hline
\textbf{$22, 151$} &$8C$ &$13A$ &$16A$ &$28A$ &$38A$ &$71A$ &$111A$ &$177A$ &$294A$ &$511A$ &$884A$ &$1518A$ &$2678A$ &$4823A$ &$8517A$ \\\hline
\textbf{$23$} &$4F$ &$7N$ &$9L$ &$15I$ &$18C$ &$31E$ &$39C$ &$65C$ &$86C$ &$158B$ &$226A$ &$375A$ &$552A$ &$930A$ &$1442A$ \\\hline
\textbf{$24, 66, 189, 231$} &$6C$ &$6S$ &$10H$ &$16M$ &$20F$ &$26F$ &$38B$ &$54D$ &$76D$ &$98B$ &$150B$ &$190A$ &$276C$ &$364A$ &$512A$ \\\hline
\textbf{$25, 61, 67, 103$} &$7D$ &$14B$ &$16B$ &$31A$ &$48B$ &$84A$ &$143A$ &$254A$ &$428A$ &$776A$ &$1355A$ &$2425A$ &$4288A$ &$7704A$ &$13740A$ \\\hline
\textbf{$26, 82, 167, 181$} &$8D$ &$11E$ &$18A$ &$31B$ &$52A$ &$84B$ &$146A$ &$263A$ &$452A$ &$790A$ &$1425A$ &$2536A$ &$4571A$ &$8224A$ &$14814A$ \\\hline
\textbf{$27, 39, 53, 83$} &$6D$ &$6T$ &$16C$ &$22B$ &$42A$ &$65A$ &$109A$ &$176A$ &$286B$ &$459A$ &$740A$ &$1209A$ &$1878A$ &$3030A$ &$4712A$ \\\hline
\textbf{$28, 70, 157, 199$} &$8E$ &$13B$ &$17A$ &$29A$ &$46A$ &$77A$ &$118A$ &$210A$ &$338A$ &$589A$ &$977A$ &$1681A$ &$2810A$ &$4967A$ &$8422A$ \\\hline
\textbf{$29, 71$} &$3A$ &$5I$ &$5C$ &$11T$ &$11X$ &$19G$ &$19I$ &$31H$ &$35A$ &$49C$ &$57B$ &$79C$ &$97B$ &$125B$ &$153A$ \\\hline
\textbf{$30, 86, 135, 149$} &$8F$ &$14C$ &$20A$ &$36A$ &$60A$ &$108A$ &$188A$ &$351A$ &$626A$ &$1161A$ &$2185A$ &$4095A$ &$7679A$ &$14594A$ &$27550A$ \\\hline
\textbf{$32, 251$} &$6E$ &$9D$ &$10I$ &$15J$ &$18D$ &$27F$ &$32E$ &$47C$ &$56C$ &$83D$ &$106E$ &$151A$ &$192A$ &$273C$ &$352B$ \\\hline
\textbf{$33, 123$} &$8G$ &$11F$ &$14J$ &$22C$ &$32A$ &$52E$ &$71B$ &$117A$ &$171A$ &$272B$ &$417B$ &$675A$ &$1071A$ &$1768A$ &$2859A$ \\\hline
\textbf{$34, 48, 187, 243$} &$4D$ &$9C$ &$6Y$ &$15H$ &$12J$ &$21D$ &$20H$ &$33F$ &$30G$ &$49B$ &$46D$ &$69D$ &$68C$ &$99B$ &$96C$ \\\hline
\textbf{$35, 49, 59, 115$} &$6F$ &$9E$ &$14K$ &$25B$ &$36B$ &$55A$ &$96A$ &$144B$ &$220A$ &$394A$ &$598A$ &$940B$ &$1621A$ &$2526A$ &$3978A$ \\\hline
\textbf{$36, 219$} &$6G$ &$6U$ &$10J$ &$12I$ &$14O$ &$20G$ &$30F$ &$42D$ &$54E$ &$86D$ &$124B$ &$168A$ &$244A$ &$350A$ &$476B$ \\\hline
\textbf{$37, 91$} &$7E$ &$13C$ &$15C$ &$26B$ &$41A$ &$69A$ &$106A$ &$180A$ &$290A$ &$508A$ &$847B$ &$1467A$ &$2600A$ &$4625A$ &$8123A$ \\\hline
\textbf{$38, 52, 155, 211$} &$6H$ &$11G$ &$11Q$ &$21A$ &$28D$ &$45D$ &$58A$ &$97A$ &$132B$ &$195B$ &$278A$ &$401A$ &$556A$ &$793A$ &$1102A$ \\\hline
\textbf{$40, 96, 235, 249$} &$7F$ &$11H$ &$15D$ &$22D$ &$30D$ &$51C$ &$75A$ &$130A$ &$193A$ &$306A$ &$502A$ &$807A$ &$1284A$ &$2096B$ &$3346A$ \\\hline
\textbf{$41, 97, 107, 121$} &$6I$ &$13D$ &$17B$ &$29B$ &$51A$ &$86A$ &$147B$ &$276A$ &$465A$ &$827A$ &$1463A$ &$2620A$ &$4630A$ &$8340A$ &$14979A$ \\\hline
\textbf{$42, 112, 171, 241$} &$5C$ &$5I$ &$7R$ &$9P$ &$13O$ &$15N$ &$19J$ &$25E$ &$29F$ &$33I$ &$43D$ &$49F$ &$59B$ &$73C$ &$85B$ \\\hline
\textbf{$43, 113$} &$3A$ &$3A$ &$7S$ &$10M$ &$16N$ &$18E$ &$39D$ &$51E$ &$96B$ &$108B$ &$241A$ &$294B$ &$605A$ &$684A$ &$1551A$ \\\hline
\textbf{$44, 100, 203, 217$} &$8H$ &$11I$ &$17C$ &$27A$ &$41B$ &$74A$ &$106B$ &$191A$ &$296A$ &$509A$ &$825A$ &$1383A$ &$2284A$ &$3812A$ &$6309A$ \\\hline
\textbf{$45, 75, 89, 101$} &$1A$ &$13E$ &$1A$ &$24B$ &$1A$ &$76A$ &$1A$ &$288A$ &$1A$ &$1135A$ &$1A$ &$3756A$ &$1A$ &$14230A$ &$1A$ \\\hline
\textbf{$46, 116, 139, 209$} &$6C$ &$9F$ &$10H$ &$15K$ &$20F$ &$29D$ &$38B$ &$55D$ &$76D$ &$103A$ &$150B$ &$195C$ &$276C$ &$371A$ &$512A$ \\\hline
\textbf{$50, 179$} &$6H$ &$12A$ &$15E$ &$24C$ &$34C$ &$56A$ &$81A$ &$131A$ &$195A$ &$329A$ &$497A$ &$830A$ &$1328A$ &$2178A$ &$3536A$ \\\hline
\textbf{$51$} &$1A$ &$1A$ &$1A$ &$1A$ &$1A$ &$1A$ &$1A$ &$1A$ &$1A$ &$1A$ &$1A$ &$1A$ &$1A$ &$1A$ &$1A$ \\\hline
\textbf{$54, 147$} &$8I$ &$13F$ &$18B$ &$27B$ &$42B$ &$67A$ &$105A$ &$173A$ &$275A$ &$463A$ &$778A$ &$1334A$ &$2282A$ &$3988A$ &$6933A$ \\\hline
\textbf{$56, 98, 185, 227$} &$7C$ &$13G$ &$19B$ &$30A$ &$45A$ &$73A$ &$106C$ &$175A$ &$255A$ &$429A$ &$647B$ &$1074A$ &$1651A$ &$2757A$ &$4283A$ \\\hline
\textbf{$57, 99$} &$4G$ &$8O$ &$10K$ &$21B$ &$26E$ &$54C$ &$71C$ &$138A$ &$209A$ &$394B$ &$632B$ &$1182A$ &$1904A$ &$3527A$ &$5857A$ \\\hline
\textbf{$58, 114, 163, 177$} &$6J$ &$14D$ &$16D$ &$33B$ &$52B$ &$90A$ &$139A$ &$254C$ &$407A$ &$728A$ &$1217A$ &$2118A$ &$3638A$ &$6395A$ &$11027A$ \\\hline
\textbf{$60, 102, 153, 195$} &$2B$ &$3C$ &$2B$ &$9Q$ &$2B$ &$3C$ &$2B$ &$5L$ &$2B$ &$3C$ &$2B$ &$17Q$ &$2B$ &$3C$ &$2B$ \\\hline
\textbf{$62, 118, 131, 145$} &$6K$ &$13H$ &$15F$ &$28B$ &$47A$ &$82A$ &$136A$ &$247A$ &$428B$ &$765A$ &$1356A$ &$2457A$ &$4408A$ &$7966A$ &$14407A$ \\\hline
\textbf{$72, 237$} &$6G$ &$9G$ &$11R$ &$15L$ &$21C$ &$31F$ &$41D$ &$59A$ &$81B$ &$115C$ &$157A$ &$219A$ &$303A$ &$421B$ &$579A$ \\\hline
\textbf{$73, 109$} &$7G$ &$12B$ &$15G$ &$24D$ &$37C$ &$62A$ &$98A$ &$162A$ &$254D$ &$437A$ &$715A$ &$1235A$ &$2095A$ &$3646A$ &$6301A$ \\\hline
\textbf{$74, 88, 173, 229$} &$6L$ &$11J$ &$16E$ &$30B$ &$52C$ &$82B$ &$147A$ &$250A$ &$445A$ &$769A$ &$1349A$ &$2376A$ &$4246A$ &$7381A$ &$13287A$ \\\hline
\textbf{$76, 205$} &$5C$ &$7O$ &$7O$ &$9R$ &$9R$ &$11Y$ &$13T$ &$15O$ &$17P$ &$19K$ &$21H$ &$25H$ &$27J$ &$33J$ &$35C$ \\\hline
\textbf{$77$} &$3B$ &$7N$ &$5J$ &$15I$ &$13P$ &$31E$ &$39E$ &$65C$ &$83C$ &$158B$ &$206B$ &$375A$ &$547A$ &$930A$ &$1411A$ \\\hline
\textbf{$78, 92, 141, 197$} &$8H$ &$14E$ &$20B$ &$34A$ &$54A$ &$91A$ &$144A$ &$254B$ &$409A$ &$720A$ &$1205A$ &$2096A$ &$3554A$ &$6224A$ &$10636A$ \\\hline
\textbf{$90, 165$} &$2B$ &$2D$ &$2B$ &$5K$ &$2B$ &$2D$ &$2B$ &$3D$ &$2B$ &$2D$ &$2B$ &$9T$ &$2B$ &$2D$ &$2B$ \\\hline
\textbf{$94, 133$} &$8J$ &$13I$ &$17D$ &$26C$ &$39B$ &$68A$ &$101A$ &$173B$ &$276B$ &$483A$ &$802A$ &$1397A$ &$2425B$ &$4288B$ &$7604A$ \\\hline
\textbf{$104, 233$} &$8K$ &$11K$ &$16F$ &$22E$ &$45B$ &$64A$ &$100A$ &$188B$ &$312A$ &$519A$ &$935A$ &$1563A$ &$2645A$ &$4799A$ &$8303A$ \\\hline
\textbf{$105$} &$1A$ &$3D$ &$1A$ &$1A$ &$2D$ &$1A$ &$1A$ &$5K$ &$1A$ &$1A$ &$2D$ &$1A$ &$1A$ &$3D$ &$1A$ \\\hline
\textbf{$106, 120, 169, 225$} &$7H$ &$10C$ &$17F$ &$30C$ &$50A$ &$95A$ &$170A$ &$286A$ &$593A$ &$1112A$ &$1958A$ &$3973A$ &$7469A$ &$13594A$ &$27091A$ \\\hline
\textbf{$108, 201$} &$7I$ &$11L$ &$13M$ &$20E$ &$28E$ &$47B$ &$63A$ &$106D$ &$162B$ &$252A$ &$396A$ &$615A$ &$979A$ &$1600A$ &$2545A$ \\\hline
\textbf{$110, 124, 137, 193$} &$6M$ &$14F$ &$20C$ &$33C$ &$52D$ &$87A$ &$146B$ &$259A$ &$434A$ &$797A$ &$1424A$ &$2583A$ &$4645A$ &$8471A$ &$15362A$ \\\hline
\textbf{$122, 161$} &$8E$ &$12C$ &$17E$ &$27C$ &$43A$ &$69B$ &$104A$ &$181A$ &$281A$ &$489A$ &$824A$ &$1422A$ &$2443A$ &$4303A$ &$7515A$ \\\hline
\textbf{$126, 129$} &$5D$ &$10D$ &$12G$ &$19F$ &$23A$ &$40A$ &$56B$ &$89A$ &$124A$ &$201A$ &$312B$ &$498A$ &$767A$ &$1254A$ &$2023A$ \\\hline
\textbf{$128, 254$} &$5E$ &$9H$ &$11S$ &$15M$ &$17K$ &$23B$ &$29E$ &$41E$ &$55E$ &$79B$ &$103B$ &$141C$ &$185A$ &$255B$ &$337A$ \\\hline
\textbf{$130, 144, 190, 246$} &$6N$ &$10E$ &$16J$ &$22F$ &$32B$ &$55B$ &$80B$ &$132A$ &$200A$ &$342A$ &$559A$ &$884B$ &$1419A$ &$2395A$ &$3893A$ \\\hline
\textbf{$132, 222$} &$6O$ &$11M$ &$14L$ &$24E$ &$30E$ &$51D$ &$77B$ &$115A$ &$164A$ &$273B$ &$421A$ &$680A$ &$1073A$ &$1742A$ &$2790A$ \\\hline
\textbf{$134, 148, 158, 214$} &$7J$ &$9I$ &$19C$ &$31C$ &$54B$ &$90B$ &$159A$ &$272A$ &$476A$ &$847A$ &$1484A$ &$2615A$ &$4681A$ &$8356A$ &$14869A$ \\\hline
\textbf{$136, 192, 238, 252$} &$8L$ &$12D$ &$16G$ &$22G$ &$33E$ &$48C$ &$68B$ &$102A$ &$146C$ &$216A$ &$315A$ &$468A$ &$683A$ &$1018A$ &$1498A$ \\\hline
\textbf{$138, 174, 208, 244$} &$5C$ &$7O$ &$7O$ &$9P$ &$13Q$ &$15N$ &$17N$ &$21G$ &$25G$ &$29G$ &$35B$ &$39F$ &$43E$ &$53D$ &$59C$ \\\hline
\textbf{$140, 196, 206, 220$} &$7K$ &$11N$ &$17G$ &$29C$ &$45C$ &$74B$ &$113A$ &$183A$ &$273A$ &$428C$ &$641A$ &$976A$ &$1445A$ &$2170A$ &$3174A$ \\\hline
\textbf{$142, 212$} &$3A$ &$3A$ &$7S$ &$10M$ &$16N$ &$18E$ &$39D$ &$51E$ &$96B$ &$108B$ &$241A$ &$294B$ &$605A$ &$684A$ &$1551A$ \\\hline
\textbf{$146, 182$} &$6P$ &$13J$ &$16H$ &$26D$ &$43B$ &$69C$ &$110A$ &$181B$ &$292A$ &$486A$ &$835A$ &$1460A$ &$2486A$ &$4359A$ &$7539A$ \\\hline
\textbf{$150$} &$1A$ &$3D$ &$1A$ &$1A$ &$2D$ &$1A$ &$1A$ &$5K$ &$1A$ &$1A$ &$2D$ &$1A$ &$1A$ &$3D$ &$1A$ \\\hline
\textbf{$152, 188, 194, 230$} &$7L$ &$14G$ &$19D$ &$31D$ &$49A$ &$91B$ &$141A$ &$253A$ &$445B$ &$798A$ &$1386A$ &$2439A$ &$4311A$ &$7662A$ &$13539A$ \\\hline
\textbf{$154, 166, 180, 210$} &$1A$ &$5C$ &$1A$ &$6AC$ &$1A$ &$5C$ &$1A$ &$7T$ &$1A$ &$5C$ &$1A$ &$10O$ &$1A$ &$5C$ &$1A$ \\\hline
\textbf{$156, 198$} &$4G$ &$8P$ &$10L$ &$22H$ &$28F$ &$57A$ &$80C$ &$150A$ &$229A$ &$445C$ &$688A$ &$1296A$ &$2102A$ &$3889A$ &$6516A$ \\\hline
\textbf{$160, 250$} &$8L$ &$9J$ &$16G$ &$17J$ &$33E$ &$27G$ &$68B$ &$53C$ &$146C$ &$87B$ &$315A$ &$176C$ &$683A$ &$334A$ &$1498A$ \\\hline
\textbf{$162, 176, 186, 242$} &$7C$ &$12E$ &$16K$ &$25C$ &$34D$ &$55C$ &$74C$ &$115B$ &$154A$ &$238A$ &$322A$ &$482A$ &$651A$ &$974A$ &$1317A$ \\\hline
\textbf{$164, 218$} &$8M$ &$10F$ &$16I$ &$27D$ &$42C$ &$61A$ &$104B$ &$176B$ &$279A$ &$447A$ &$768A$ &$1327A$ &$2099A$ &$4048A$ &$6717A$ \\\hline
\textbf{$168, 224, 234, 248$} &$6Q$ &$10G$ &$16L$ &$27E$ &$44A$ &$76B$ &$125A$ &$217A$ &$365A$ &$632A$ &$1069A$ &$1840A$ &$3120A$ &$5365A$ &$9109A$ \\\hline
\textbf{$170, 240$} &$1A$ &$1A$ &$1A$ &$1A$ &$1A$ &$1A$ &$1A$ &$1A$ &$1A$ &$1A$ &$1A$ &$1A$ &$1A$ &$1A$ &$1A$ \\\hline
\textbf{$172, 202, 216, 228$} &$7K$ &$11O$ &$13N$ &$28C$ &$41C$ &$73B$ &$120A$ &$206A$ &$335A$ &$573A$ &$951A$ &$1596A$ &$2632A$ &$4394A$ &$7238A$ \\\hline
\textbf{$178$} &$3B$ &$7N$ &$5J$ &$15I$ &$13P$ &$31E$ &$39E$ &$65C$ &$83C$ &$158B$ &$206B$ &$375A$ &$547A$ &$930A$ &$1411A$ \\\hline
\textbf{$184, 226$} &$3A$ &$6V$ &$7S$ &$14M$ &$16N$ &$32D$ &$39D$ &$76C$ &$96B$ &$186A$ &$241A$ &$461A$ &$605A$ &$1169A$ &$1551A$ \\\hline
\textbf{$200, 236$} &$5F$ &$7P$ &$9M$ &$11V$ &$13R$ &$17L$ &$21E$ &$27I$ &$33G$ &$41F$ &$49D$ &$61B$ &$75B$ &$93A$ &$113B$ \\\hline
\textbf{$204$} &$1A$ &$1A$ &$1A$ &$1A$ &$1A$ &$1A$ &$1A$ &$1A$ &$1A$ &$1A$ &$1A$ &$1A$ &$1A$ &$1A$ &$1A$ \\\hline
\textbf{$232$} &$4F$ &$7N$ &$9L$ &$15I$ &$18C$ &$31E$ &$39C$ &$65C$ &$86C$ &$158B$ &$226A$ &$375A$ &$552A$ &$930A$ &$1442A$ \\
\label{longt}
\end{longtable}
}}
\end{landscape}

\end{document}